\documentclass[twocolumn,showpacs,aps]{revtex4}

\usepackage{graphicx}
\usepackage{amssymb}

\begin{document}

\title{Anharmonic vs. relaxational sound damping in glasses:\\	
II. Vitreous silica}

\author{Ren\'e Vacher, Eric Courtens, and Marie Foret}

\affiliation{Groupe de Physique des Verres et Spectroscopies, LCVN, UMR CNRS 5587,\\
Universit\'e Montpellier II, F-34095 Montpellier Cedex 5, France}

\date{\today}

\begin{abstract}
The temperature dependence of the frequency dispersion in the sound velocity and damping
of vitreous silica is reanalyzed.
Thermally activated relaxation accounts for the sound attenuation observed above 10 K at
{\em sonic} and {\em ultrasonic} frequencies.
Its extrapolation to the {\em hypersonic} regime reveals that the anharmonic coupling to the thermal bath becomes important
in Brillouin-scattering measurements.
At 35 GHz and room temperature, the damping due to this anharmonicity is found to be nearly twice that produced by thermally
activated relaxation.
The analysis also reveals a sizeable velocity increase with temperature which is not
related with sound dispersion.
A possible explanation is that silica experiences a gradual structural change that already
starts well below room temperature.
\end{abstract}


\maketitle

\section{Introduction}

In the first paper of this series (I), the hypersonic attenuation of highly densified
silica glass, $d$-SiO$_2$, was investigated.
It was found that in that material the damping of hypersound is completely
dominated by its {\em anharmonic interaction} with the thermally dominant modes.
In the present paper we consider the corresponding situation in usual vitreous silica, $v$-SiO$_2$,
for which a large quantity of high quality data is already available in the literature.
In the early 1950's, ultrasonic absorption peaks in
function of the temperature $T$ were first observed in $v$-SiO$_2$.
These were described by Anderson and B\"ommel in terms of a phenomenological Maxwell model \cite{And55}.
Following the discovery of the key role played by two-level systems (TLS) in producing the anomalous
thermal properties of glasses \cite{And72,Phi72}, it became clear that these should
also be invoked in the description of acoustic relaxation.
A theory including both resonant and relaxational sound damping produced by tunneling was
developed by J\"ackle \cite{Jae72}.
It was then extended to higher temperatures by including in the description the classical jumps over
the energy barrier separating the wells of TLS \cite{Jae76}.
The presently accepted model for thermally activated relaxation (TAR) includes a distribution of asymmetric double-well potentials,
as discussed by Gilroy and Phillips \cite{Gil81}.

The simple extrapolation of this behavior to GHz frequencies, {\em i.e.} to the hypersonic regime,
has sometimes been presented as an appropriate description of the observed phenomena, {\em e.g.} in 
\cite{Wie00,Wie01} or \cite{Sur04}.
However, it has also been recognized that TAR might not always be sufficient to account for sound
dispersion and damping.
A specific example has been discussed in the case of vitreous germania, GeO$_2$ \cite{Her98}.
The results presented in (I) strongly suggest that $v$-SiO$_2$ should be an excellent candidate to search for a
possible anharmonic contribution to the hypersonic attenuation.

The paper is organized as follows.
In Section II the phenomenology of TAR is reviewed to formulate a description that can reasonably be
extrapolated to sufficiently high-$T$ and to Brillouin-scattering frequencies.
In Section III this formalism is applied to available high quality sonic and ultrasonic data on $v$-SiO$_2$,
covering more than four orders of magnitude in the frequency $\nu$.
The model parameters that are thus extracted are then used in Section IV to estimate the TAR contributions
to both velocity dispersion and damping at Brillouin-scattering frequencies.
The anharmonic contribution can then be extracted from the total signal.
It is analyzed in terms of a mean thermal mode relaxation time $\tau_{\rm th}$ in Section V.
Interestingly, this time is found to be about an order of magnitude longer than that of densified silica glass, $d$-SiO$_2$.
Section VI presents a synthesis of the above analysis for the entire range of $\nu$,
from ultrasonic to Brillouin-scattering frequencies.
A part of that analysis concerns the velocity changes with $T$.
We find an anomalous increase of the bare velocity $v_\infty$ with $T$, nearly the same in
$v$-SiO$_2$ and $d$-SiO$_2$.
This unrelaxed velocity, $v_\infty$, is obtained after subtraction of the velocity changes produced by both
thermally activated relaxation and anharmonicity.
The $T$-dependence of $v_\infty$ suggests that silica experiences a progressive structural change with increasing $T$.
Section VII is a discussion, which mentions the crossovers between the various frequency regimes,
considers the suitability of power laws for the description
of the dependence of the damping on $\nu$, and evoques possible extension to other glasses.

\section{The phenomenology of thermally activated relaxation}

We consider an assembly of defects represented by double-well potentials that are separated by barriers of height $V$
and whose depths differ by the asymmetry $\Delta$.
The energies $V$ and $\Delta$ are randomly distributed according to a distribution $P(\Delta , V) \: d\Delta \: dV$ to be discussed
below.
The system is thought to hop continuously between the wells.
The energy difference between the wells is coupled to the strain $\bf e$ of a sound wave of angular frequency
$\Omega \equiv 2 \pi \, \nu$ by a deformation potential
$\gamma  = {\scriptstyle \frac{1}{2}} \partial \Delta / \partial e $.
Owing to the delayed energy exchanged in hopping, this produces the dissipation of the sound wave.
This situation has been described in great details elsewhere \cite{Jae76,Gil81,Phi87,Tie92}.
To an excellent approximation, it leads to a relaxational contribution to the internal friction given by
\begin{widetext}
$$Q_{\rm rel}^{-1} \; = \; \frac {\gamma ^2}{\rho v^2 T} \; \int_{-\infty}^{\infty} d\Delta \; \int_{0}^{\infty}  dV \; P(\Delta , V) \; 
{\rm sech}^2 \frac {\Delta}{2T} \; \frac {\Omega \tau}{1 + \Omega ^2 \tau ^2} \; \; . \eqno{(1a)}$$
In this expression, $\rho$ is the material density, $v$ is the velocity of sound, and $T$ is in energy units.
We remark that both $\gamma$ and $v$ depend on the polarization of the acoustic wave, longitudinal (LA) or transverse (TA).
In this writing, $V$ is restricted to positive values while the distribution is symmetric in $\Delta$.
We also note that $P(\Delta , V) \: d\Delta \: dV$ is a density, {\em i.e.} a number per unit volume.
The associated velocity change, which follows from the Kramers-Kronig relation, is given by
$$\left( \frac {\delta v}{v} \right)_{\rm rel} \; = \; -\frac {1}{2} \; \frac {\gamma ^2}{\rho v^2 T} \;
\int_{-\infty}^{\infty} d\Delta \;  \int_{0}^{\infty}  dV \; P(\Delta , V) \; 
{\rm sech}^2 \frac {\Delta}{2T} \; \frac {1}{1 + \Omega ^2 \tau ^2} \; \;. \eqno{(1b)}$$
\end{widetext}
In these equations, $\tau$ is the relaxation time for hopping within the double well.
It is given by
$$\tau \; = \; \tau _0 \; {\rm exp} \, \frac{V}{T} \; {\rm sech} \, \frac {\Delta}{2T} \; \; , \eqno{(2)}$$
where $\tau _0$ is the inverse of an attempt frequency, as shown in detail {\em e.g.} in \cite{Tie92}.

The key in applying these expressions is to use a reasonable distribution $P(\Delta , V)$.
For small $\Delta$ and $V$ the distribution is often replaced by a constant $\bar{P}$.
This is suggested by the $T$-dependence of the specific heat at low $T$ which only probes
low values of $\Delta$ and $V$ \cite{Poh81}.
Of course, a constant $\bar{P}$ cannot be extended to high values of $\Delta$ and $V$ as this
leads to a diverging integral density of defects which is unphysical.
A reasonable guess for $P(\Delta , V)$ can be obtained with the help of the soft-potential model (SPM) \cite{Kar83}.
That model is characterized by a distribution
of random dimensionless cubic and quadratic coefficients, $\xi$ and $\eta$ respectively, by an energy scale of
the potential ${\cal E}_0$, and by a characteristic cross-over energy $W \ll {\cal E}_0$ \cite{Kar83}.
We are only interested here in the region $\eta < 0$ with $1 \gg |\eta | \gg \xi ^2$ which gives double wells with
barriers centered at the origin of the soft mode coordinate $x$.
As shown in \cite{Ili87}, owing to the latter choice, the variables $\xi$ and $\eta$ are not statistically independent.
This leads
to a sea-gull singularity in their distribution, $P(\xi, \eta) = {\scriptstyle \frac {1}{2}} |\eta | {\cal P}_0(\xi, \eta)$,
where ${\cal P}_0(\xi, \eta)$ is finite near the origin.
For the range of values of interest here, one has $\Delta = {\cal E}_0 {\scriptstyle \frac {1}{\sqrt{2}}} \xi |\eta |^{3/2}$
and $V = {\scriptstyle \frac {1}{4}} {\cal E}_0 |\eta |^2$.
The deformation potential of the SPM is also function of $|\eta |$ with
$\gamma_{\rm SPM}^2 \propto |\eta |$ as defined in \cite{Buc92} and further explained in \cite{Par94}.
The terms $\gamma^2 P(\Delta , V) \, d\Delta \, dV$ of (1$a$) are transformed into
$\gamma_{\rm SPM}^2 P(\xi, \eta) \, d\xi \, d\eta$ in the SPM.
Using the Jacobian $|\partial (\Delta ,V)/\partial (\xi, \eta) | = {\cal E}_0^2 
{\scriptstyle \frac {1}{2\sqrt{2}}} |\eta |^{5/2}$, one finds that
$P(\Delta , V) \propto V^{-1/4} {\cal P}_0(\xi, \eta)$ \cite{Buc92}.
This was already used by Keil {\em et al.} \cite{Kei93} who selected a distribution $P(V)$
proportional to $V^{-\zeta}$ times a gaussian.
These authors experimentally found that indeed $\zeta$ is very near 1/4 in the case of silica.

For convenience, and for lack of different compelling indications, we assume that the distribution can be factored into
$$P(\Delta , V) \; = \; f(\Delta) \; g(V) \; \; . \eqno{(3)}$$
For $g(V)$, inspired by \cite{Buc92} and \cite{Kei93}, we use the normalized form
$$g(V) \; = \; N_{\rm g} \; \frac{1}{V_0} \; \left( \frac{V}{V_0} \right) ^{- \zeta} {\rm exp} 
\left( - \frac{1}{2} \, \frac {V^2}{V_0^2} \right) \; \; . \eqno{(4a)}$$
Since $\zeta < 1$, this expression is integrable.
The norm $N_{\rm g}$ is selected so that $g(V)$ integrates to 1. This gives
$$N_{\rm g}^{-1} \; = \; \int_{0}^{\infty} x^{- \zeta} {\rm exp} ( - {\scriptstyle \frac{1}{2}} x^2 ) dx \; 
= \; \Gamma ( 1 - \zeta) \; U({\scriptstyle \frac{1}{2}} - \zeta , 0) \; \; . \eqno{(4b)}$$
The function $U(a,0)$ is the parabolic cylinder function \cite{Abr70}.
As explained in the following Section, our
independent analysis of a collection of data larger than that used in \cite{Kei93} also leads
to $\zeta$ very close to 1/4. 
This provides a solid support for this particular choice of $g(V)$.
For $f(\Delta)$ we use a simple gaussian, 
rather similar to the gaussian cut-off of the linear asymmetry coefficient $D_1$ used by Gil {\em et al.} \cite{Gil93}.
Indeed, $D_1 \propto \xi |\eta | \propto \Delta V^{-1/4}$,
so that a gaussian in $D_1$ is very close to one in $\Delta$, the power of $V$ connecting the two variables being quite small.
This functional form for $f(\Delta)$ was already employed succesfully by Bonnet \cite{Bon91}. It is written
$$f(\Delta) \; = \; N_{\rm f} \: f_0 \: {\rm exp} \left( - \frac{1}{2} \, \frac{\Delta ^2}{\Delta _{\rm C}^2} \right) \; \; , \eqno{(5a)}$$
where $f_0$ is defined by the normalization condition that the integral of $f(\Delta )$
equals $f_0 V_0$. The norm $N_{\rm f}$ is then dimensionless and given by
$$N_{\rm f}^{-1} \; = \; \int_{-\infty}^{\infty} {\rm exp} ( - {\scriptstyle \frac{1}{2}} \delta ^2 x^2) \, dx \;
= \; \sqrt{2\pi} / \delta \; \; , \eqno{(5b)}$$
where we defined
$$\delta \; \equiv \; V_0 / \Delta _{\rm C} \; \; , \eqno{(6)}$$
$\Delta _{\rm C}$ being the cut-off value of the asymmetry.
We finally remark that with the above definitions, the integral density of defects is
$$N \; \equiv \; \int_{-\infty}^{\infty} d\Delta \; \int_{0}^{\infty}  dV \; P(\Delta , V) \; = \; f_0 V_0 \; \; . \eqno{(7)}$$

To complete the calculation, we now introduce (4) and (5) into (1), and use (2).
In performing the integrals in (1), we make the same approximation as in \cite{Gil81} that
sech($\Delta / 2T$) is replaced by 1 for $|\Delta |\; < \; 2T$ and by zero otherwise.
This eliminates the sech factors in (1) and (2) and simply
replaces the limits of integration on $\Delta$ by $\pm 2T$.
It is convenient to define a dimensionless constant
$${\cal C} \; \equiv \; \gamma^2 f_0 N_{\rm g} / \rho v^2 \; \; . \eqno{(8)}$$
One obtains
\begin{widetext}
$$Q_{\rm rel}^{-1} \; = \; {\cal C} \: \Phi \left( \frac{\sqrt{2} \, T}{ \Delta _{\rm C} } \right) \: \frac {1}{T} \: \int_{0}^{\infty}
\left( \frac{V}{V_0} \right) ^{- \zeta} {\rm exp} \left( - \frac{1}{2} \, \frac {V^2}{V_0^2} \right)
\; \frac {\Omega \tau _0 \: {\rm exp}(V/T)}{1 + \Omega ^2 \tau_0 ^2 \: {\rm exp} (2V/T)} \: dV \; \; , \eqno{(9a)}$$
$$\left( \frac {\delta v}{v} \right)_{\rm rel} \; = \; -\frac {1}{2} \; {\cal C} \: \Phi \left( \frac{\sqrt{2} \, T}{\Delta _{\rm C}} \right) 
\: \frac {1}{T} \: \int_{0}^{\infty}
\left( \frac{V}{V_0} \right) ^{- \zeta} {\rm exp} \left( - \frac{1}{2} \, \frac {V^2}{V_0^2} \right)
\; \frac {1}{1 + \Omega ^2 \tau_0 ^2 \: {\rm exp} (2V/T)} \: dV \; \; , \eqno{(9b)}$$
\end{widetext}
where $\Phi (z)$ is the error function,
$$\Phi (z) \; \equiv \; \frac{2}{\sqrt{\pi}} \; \int_{0}^{z} {\rm exp} (-x^2) dx \; \; . \eqno{(10)}$$
For the purpose of comparison with literature results, it will be useful to relate the value of $\cal{C}$
with the ``tunneling strength'' defined by $C \equiv \bar{P} \gamma^2/\rho v^2$ in the usual
tunneling model  \cite{Phi87,Buc92,Poh02}.
Given the distribution (4), one cannot strictly define $\bar{P}$, but this does not prevent defining
$C$ in a consistent manner \cite{Buc92}.
We calculate this for $\zeta = {\scriptstyle \frac{1}{4}}$.
We use $\gamma_{\rm SPM}^2 = {\scriptstyle \frac{1}{2}} \Lambda ^2 |\eta |$ and
$C = \Lambda ^2 {\cal P}_0 \, \eta_{\rm L}^{7/2} \, / \,W \rho v^2$
from \cite{Buc92}, where $\eta_{\rm L}^2 \equiv W/{\cal E}_0$.
Introducing these definitions into 
$\gamma_{\rm SPM}^2 P(\xi, \eta) \, d\xi \, d\eta \, = \, \gamma^2 P(\Delta , V) \, d\Delta \, dV$, and using (8),
we find in the limit of small $\Delta$ and $V$,
$$ C \; = \; \sqrt{\frac{2}{\pi}} \; \frac{W^{3/4}V_0^{1/4}}{\Delta _{\rm C}} \; {\cal C} \; \; . \eqno{(11)}$$
Information on the numerical handling of Eqs. (9) and our choice of suitable fitting parameters are
found in Appendix A.

\section{Analysis of sonic and ultrasonic relaxational data}

It is generally agreed that sonic or ultrasonic damping at temperatures above $\sim 10$ K is dominated
by thermally activated relaxation.
To obtain the model parameters $V_0$, $\Delta _{\rm C}$, and $\tau_0$ entering Eqs.$\; (9)$, it is necessary to
analyze acoustic results over a sufficiently large range of frequencies $\nu$, this up to high ultrasonic frequencies.
It implies comparing data from various sources and generally acquired with different measurement techniques.
TAR leads to a peak in $Q_{\rm rel}^{-1}$ as observed by Anderson and B\"ommel in ultrasonic
pulse-echo measurements \cite{And55}.
These authors report precise data obtained on the TA-mode of vitreous silica at $\nu =20$ MHz.
This is one of the curves that will be used for our analysis, as displayed in Fig.$\; 1$.
Other high-frequency data were measured using the Bragg diffraction of light, \cite{Vac81} in a
set-up similar to the original Debye-Sears experiment \cite{Deb32}.
Damping results at $\nu \simeq 200$ MHz on both LA- and TA-modes have been obtained in that manner \cite{Vac81}.
From these, and from results presented in \cite{Str64}, it is clear that LA- and TA-modes lead to identical $Q^{-1}$ peak shapes.
This implies that similar distributions of defects are active in the damping of all acoustic waves,
independently from their polarizations.
We use the data on the LA-mode at 207 MHz from \cite{Vac81}, also shown in Fig.$\: 1$.
The independence from polarization allows including in the evaluation sonic frequency results acquired on macroscopic
vibrational modes, such as in vibrating reed measurements.
We use the data at $\nu =11.4$ kHz of Classen, as reported in \cite{Tie92}.
This curve is also displayed in Fig.$\: 1$.
Results at intermediate frequencies were collected using composite oscillators, at 660 kHz on the LA-mode
in \cite{Gil81}, and at 180 kHz on a torsional mode in \cite{Kei93}.
As remarked in \cite{Kei93}, this particular method can easily lead to instrumental background-loss contributions.
This might have been the case in \cite{Gil81}, as suggested from data on other glasses
presented by the same authors in \cite{Gil80} which show long absorption tails at high $T$.
This notion is also supported by fits that are explained below.
Hence, it is the 180 kHz data from \cite{Kei93} which is included in Fig.$\; 1$.
With these four curves, the analysed data covers more than four decades in $\nu$ with a nearly
linear progression in ${\rm log} \, \nu$.

\begin{figure}
\includegraphics[width=8.5cm]{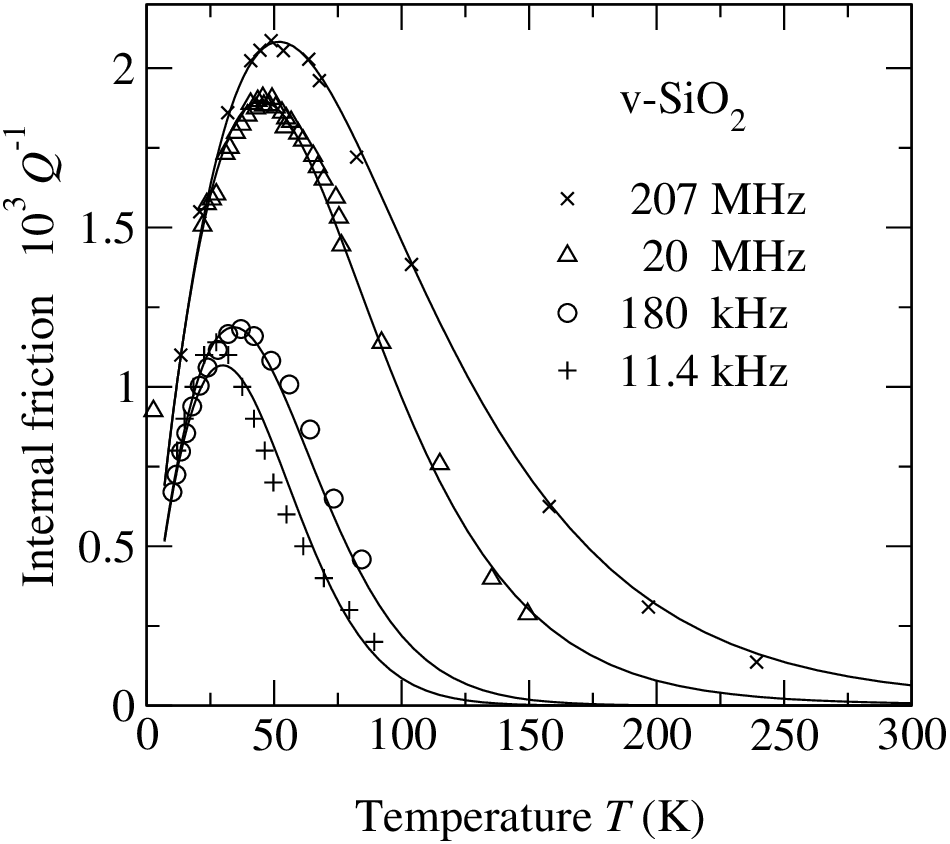}
\caption{The data points are internal friction results on $v$-SiO$_2$ taken from the literature, 
\cite{Vac81}, \cite{And55}, \cite{Kei93}, and \cite{Tie92} in order of decreasing frequency.
The solid lines show our adjustment to Eq.$\; (9a)$ as explained below.}
\end{figure}
The curves shown in Fig.$\; 1$ have been simultaneously adjusted to Eq.$\; (9a)$, allowing for each
measurement an independent coupling parameter $\cal C$.
In the Marquardt-Levenberg routine used to this effect, the weights have been adjusted so that each
curve had approximately the same weight, independently from the number of data points
at each particular frequency.
The excellent results of this fit, with $\zeta$ fixed at 1/4, are illustrated by the solid lines in Fig.$\; 1$.
The distribution parameters corresponding to these lines are $V_0 = 659 \pm 19$ K, ${\rm log}_{10} \tau_0 = -12.2 \pm 0.14$
with $\tau_0$ in seconds, and $V_0/ \Delta _{\rm C} = 8.2 \pm 0.6$.
If the same fit is repeated with $\zeta$ free, one finds $\zeta = 0.28 \pm 0.03$, and similar values for the other
parameters: $V_0 = 667 \pm 21$ K, ${\rm log}_{10} \tau_0 = -12.2 \pm 0.18$, and $V_0/ \Delta _{\rm C} = 7.7 \pm 0.7$.
This shows that the distribution $g(V)$ suggested by the SPM is very adequate indeed.
It also shows that $\zeta$ can be fixed at 1/4, which will be done for the rest of this paper.
The rather high value obtained for $V_0/ \Delta _{\rm C}$ emphasizes that
one should not neglect the cut-off in the asymmetry distribution.
We return to this point below.

Our analysis reveals a certain difficulty in comparing the absolute size of the various curves.
This can be related to some extent to the different polarizations of the waves in these measurements, or
also to a small part to the different qualities of the silica samples employed.
However, it seems more likely that it arises to a large part 
from calibration inaccuracies in some of these measurements.
Indeed, discrepancies in the size of $Q^{-1}$ are directly observed in comparing literature reports,
for example for the TA-waves at 20 MHz in \cite{And55} and \cite{Str64}.
The difficulty is also seen in comparing the LA to TA values of $Q^{-1}$ in \cite{Vac81} and in \cite{Str64}.
While in the former case it is the attenuations that have the same amplitudes for both LA and TA, in the latter it is
the internal frictions which are practically the same for both polarizations.
This cannot be reconciled as attenuation and friction differ by a factor proportional to the sound velocity, and
the latter is $\simeq 5900$ m/s for LA and $\simeq 3800$ m/s for TA-waves at low $T$.
We also observed that the results at 660 kHz in \cite{Gil81} are fitted extremely well with the same model parameters,
provided one allows for a relatively small constant background contribution to $Q^{-1}$, equal to $0.1 \times 10^{-3}$.
The values of $\cal C$ obtained for these various measurements range then from $1.1 \times 10^{-3}$ in the
case of \cite{Gil81} to $1.8 \times 10^{-3}$ for \cite{And55}. These variations in $\cal C$ do not seem
correlated to either the measuring frequency or the wave polarization, which is another reason to
suspect calibration difficulties.

\begin{figure}
\includegraphics[width=8.5cm]{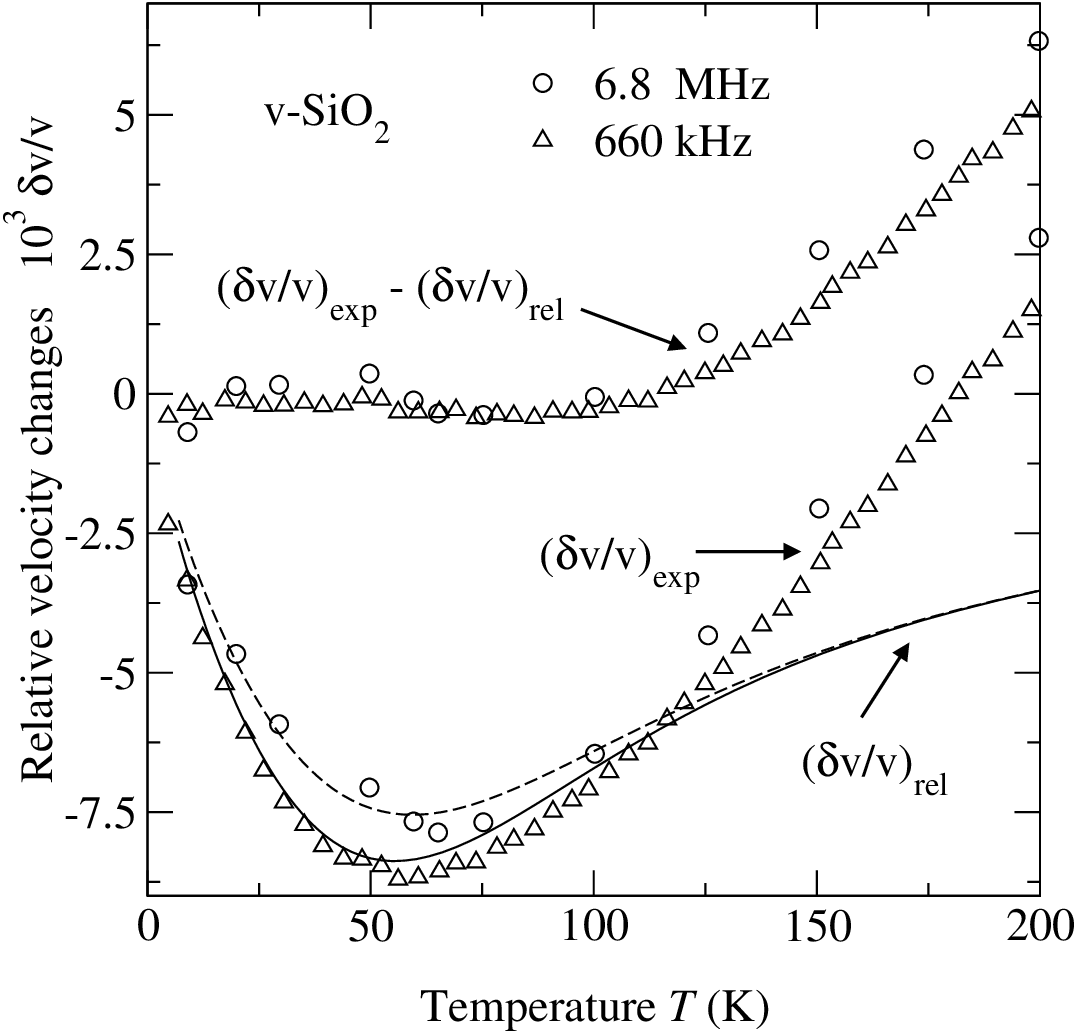}
\caption{The data points of the two lower curves are relative velocity changes of $v$-SiO$_2$ taken from the literature, \cite{Gil81} for 660 kHz and
\cite{Vac81} for 6.8 MHz.
The solid lines show our calculation of the TAR contribution according to Eq.$\; (9b)$ with ${\cal C} = 1.4 \times 10^{-3}$.
The data points of the upper two curves show the experiment minus the TAR contribution.}
\end{figure}
It is necessary to obtain a more reliable value of $\cal C$ to analyze the Brillouin scattering data in the
following Section.
To this effect we remark that velocity measurements are generally both accurate and precise, allowing to follow
small changes in $\delta v / v$ with excellent accuracy.
We use two measurements of the LA-wave velocity, one at 660 kHz from \cite{Gil81} and the other at 6.8 MHz from \cite{Vac81}.
The experimental values $(\delta v / v)_{\rm exp}$ are shown in Fig.$\: 2$, together with continuous lines marked
$(\delta v / v)_{\rm rel}$.
These are  calculated from the relaxation equation $(9b)$ using the
above values of $V_0$, $\Delta _{\rm C}$, and $\tau_0$, together with ${\cal C} = 1.4 \times 10^{-3}$.
With the latter value, the dip around 50 K is completely accounted for.
This is emphasized by the difference $(\delta v / v)_{\rm exp} - (\delta v / v)_{\rm rel}$ also shown in Fig.$\; 2$.
Hence, we adopt for the following the coupling constant ${\cal C} = 1.4 \times 10^{-3}$ which happens to fall
within the range of values obtained by fitting the various $Q^{-1}$ curves discussed above.

We now return to the small cut-off value $\Delta_{\rm C} \simeq 80$~K which we obtained.
This aspect has mostly been ignored by other workers.
It was shown in \cite{Bon91} that it is necessary to include a cut-off to obtain a good fit to the high-$T$ tails
in Fig.~1.
For example, forcing $\Delta_{\rm C} = V_0$, the quality of the fits degrades appreciably above $\sim 70$~K,
especially for the highest $\nu$ curve.
More importantly, the calculated values of $(\delta v / v)_{\rm rel}$ are then much too large to account
properly for the dip around 60~K observed in Fig.~2.
Obviously, by the virtue of the Kramers-Kronig transform, $(\delta v / v)_{\rm rel}$ integrates over a
large part of the distribution, which is the reason for this problem.
One might wish to gain an intuitive picture of why $\Delta_{\rm C}$ can be so much smaller than $V_0$.
To this effect, one can consider the models of TAR drawn in Fig.~2 of \cite{Hun76}.
We take as simplest examples model A, in which the two Si atoms of a Si$-$O$-$Si bond are too close, and
model B, in which they are too far apart.
In either case, a double well potential for the connecting oxygen results with a barrier height $V$
that comes mainly from the separation of the two Si.
This barrier can thus be quite high.
On the other hand, the asymmetry is produced by the difference in the wider environment of the two wells.
In a hard glass, one might expect that these environments, which are dictated by the minimization of the energy, can be
mostly quite similar.
From these considerations, one intuitively anticipates that the ratio $V_0/\Delta_{\rm C}$ might depend
significantly on the particular glass.

Finally, we remark that the temperature $T_{\rm max}$ of the peak positions in Fig.$\; 1$ depends linearly on ln$\, \Omega$.
A similar observation was already made in \cite{Hun76}.
This typical Arrhenius behavior supports TAR as the principal relaxation mechanism to describe Fig.$\; 1$, as opposed
for example to incoherent tunneling that would lead to $T_{\rm max} \propto \sqrt{\Omega}$ as seen from Eq. (2.95) of \cite{Rau95}.

\section{Analysis of Brillouin scattering results}

High-resolution Brillouin scattering measurements of the temperature dependence of the LA-linewidth of vitreous silica have been reported
in \cite{Vac76} and \cite{Tie92}.
Both experiments were performed near backscattering and below room $T$, using as exciting radiation the blue argon-laser
line at $\lambda_{\rm L} = 488$ nm.
Measurements above room $T$ are reported in \cite{Pel76}.
These were performed at $\lambda_{\rm L} = 514$ nm.
The Brillouin frequency shifts $\nu_{\rm B}$ from \cite{Vac76} and \cite{Pel76}, the latter rescaled to $\lambda_{\rm L} = 488$ nm,
are displayed in Fig.$\; (3a)$ over the entire range of $T$.
\begin{figure}
\includegraphics[width=8.5cm]{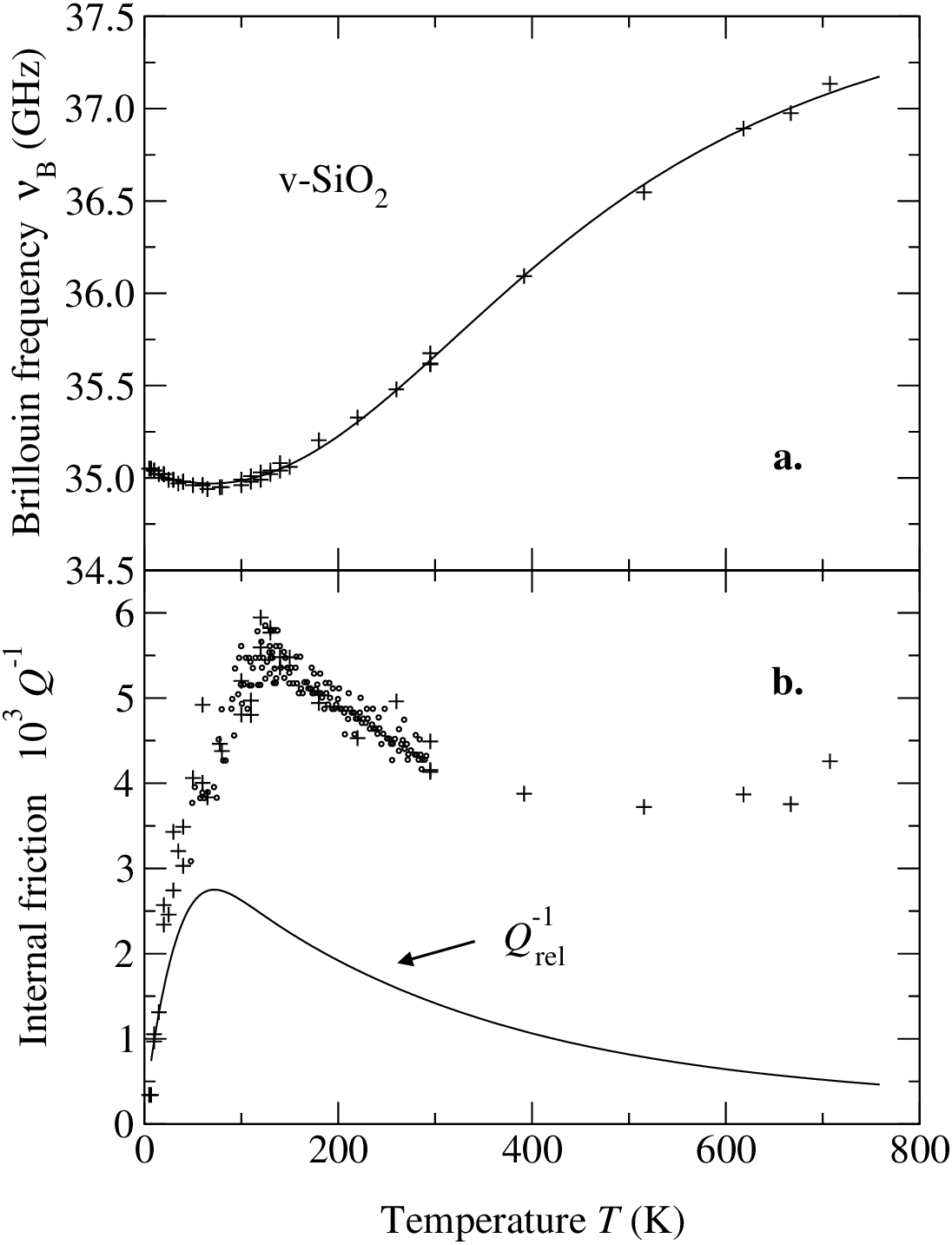}
\caption{{\bf a}: The data points are Brillouin shifts obtained on $v$-SiO$_2$ near backscattering,
from \cite{Vac76} and \cite{Pel76}. The values of \cite{Pel76} have been
rescaled to the blue argon-laser line, $\lambda_{\rm L} = 488$ nm.
The solid line is a guide to the eye.
{\bf b}: The internal friction derived from the Brillouin linewidths. The crosses are from \cite{Vac76} and \cite{Pel76}, 
while the dots are from \cite{Tie92}. The solid curve shows the TAR contribution calculated from Eq.$\; (9a)$ using the value
of $\cal C$ determined from the velocities in Fig. 2.}
\end{figure}

The frequency widths extracted from the Brillouin peaks in \cite{Vac76} and \cite{Pel76},
after correction for the instrumental broadening,
have been converted to internal friction as shown in Fig.$\; (3b)$.
The internal friction reported by Tielb\"urger {\em et al}. in \cite{Tie92} is also shown on the same figure.
One notices the remarkable agreement between these two independent determinations.
It emphasizes that Brillouin scattering gives an absolute measurement of $Q^{-1}$, independent from
calibrations, since it suffices to take the ratio of the Brillouin full-width to the Brillouin shift to
extract $Q^{-1}$.
This statement of course assumes that the spectroscopy can be performed with the required resolution, 
and in particular that the broadening produced by the finite collection aperture can be made sufficiently small.

Also shown in Fig.$\; (3b)$ is the curve $Q_{\rm rel}^{-1}$ calculated with Eq.$\; (9a)$ for $\nu = 35$ GHz and 
with the parameters determined in Section III.
It is evident that at Brillouin frequencies TAR cannot account for the entire internal friction.
The value ${\cal C} \simeq 1.4 \times 10^{-3}$ is confirmed by the Brillouin velocity results explained below.
However, even if one would allow oneself to arbitrarily double the coupling constant $\cal C$, the line in Fig.$\; 3(b)$
would still not superpose the data points.
In particular the peak in $Q_{\rm rel}^{-1}$ occurs at a lower $T$-value than the peak in the observed $Q^{-1}$.
The difference between the two curves is attributed to its largest part to the anharmonic coupling of 
the LA-waves with the thermally excited vibrational modes, as reported in (I) for densified silica glass.

The velocities are extracted from the Brillouin frequencies $\nu_{\rm B}$ of Fig.$\; (3a)$ using
$v = \lambda_{\rm L} \nu_{\rm B}\, / \, 2 n \, {\rm sin} \, {\frac{\theta}{2}}$, where $n$ is the 
refractive index and $\theta$ is the internal scattering angle, here close to $180^{\circ}$.
To derive precise values on $v(T)$, it is necessary to know $n(T)$ with an equal precision.
This information has been derived from \cite{Wax71}.
The results below 300 K are shown in Fig.~4, together with the velocities at 6.8~MHz from \cite{Vac81}.
\begin{figure}
\includegraphics[width=8.5cm]{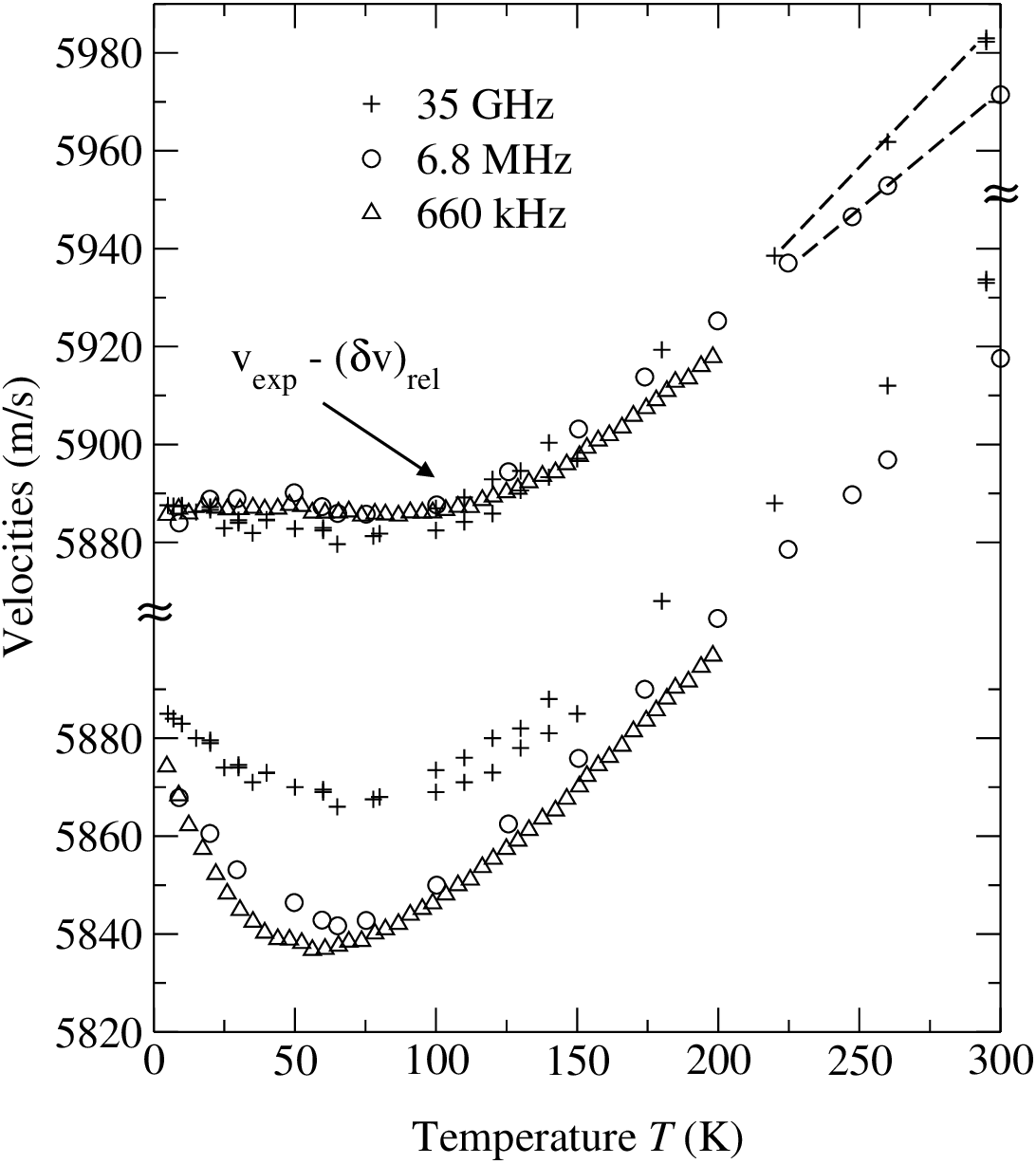}
\caption{The lower data points show the measured ultrasonic velocities from Fig.~2 together with the hypersonic velocity
derived from Fig.$\; (3a)$. The upper data points show the same velocitites after subtraction of the TAR contribution.
We note that after this subtraction there remains a small difference between the slopes observed at high $T$ for the two
different frequencies. These slopes are shown by the dashed lines and they are discussed in Appendix B.}
\end{figure}
The relative changes reported in \cite{Gil81} have also been converted back to velocities using $v_0 = 5888$~m/s.
Subtracting from all three curves the $(\delta v)_{\rm rel}$ calculated from Eq.$\; (9b)$, one
observes that below 150~K all the data collapse quite well.
This confirms that the value of $\cal C$ is also correct at Brillouin frequencies.
It also emphasizes that the $\nu$-dependence in the depth of the dip around 70~K is well predicted by TAR.
As explained in Section VIB, the velocity changes that are produced by the anharmonicity, $(\delta v)_{\rm anh}$, show
little dispersion compared to the large dispersion in $(\delta v)_{\rm rel}$.
Hence, they do not modify the above conclusion.

Figure~5 shows the difference between the experimental internal friction and the calculated TAR contribution.
Similarly to the results on $d$-SiO$_2$ in (I), there is a region where the signal falls rapidly with decreasing $T$.
However, it does not seem to fall to zero sufficiently fast.
In spite of the large scatter in the experimental values, there is a hint for an additional contribution to $Q^{-1}$
at low $T$, in the region from $\sim 20$ to $\sim 60$ K.
\begin{figure}
\includegraphics[width=8.5cm]{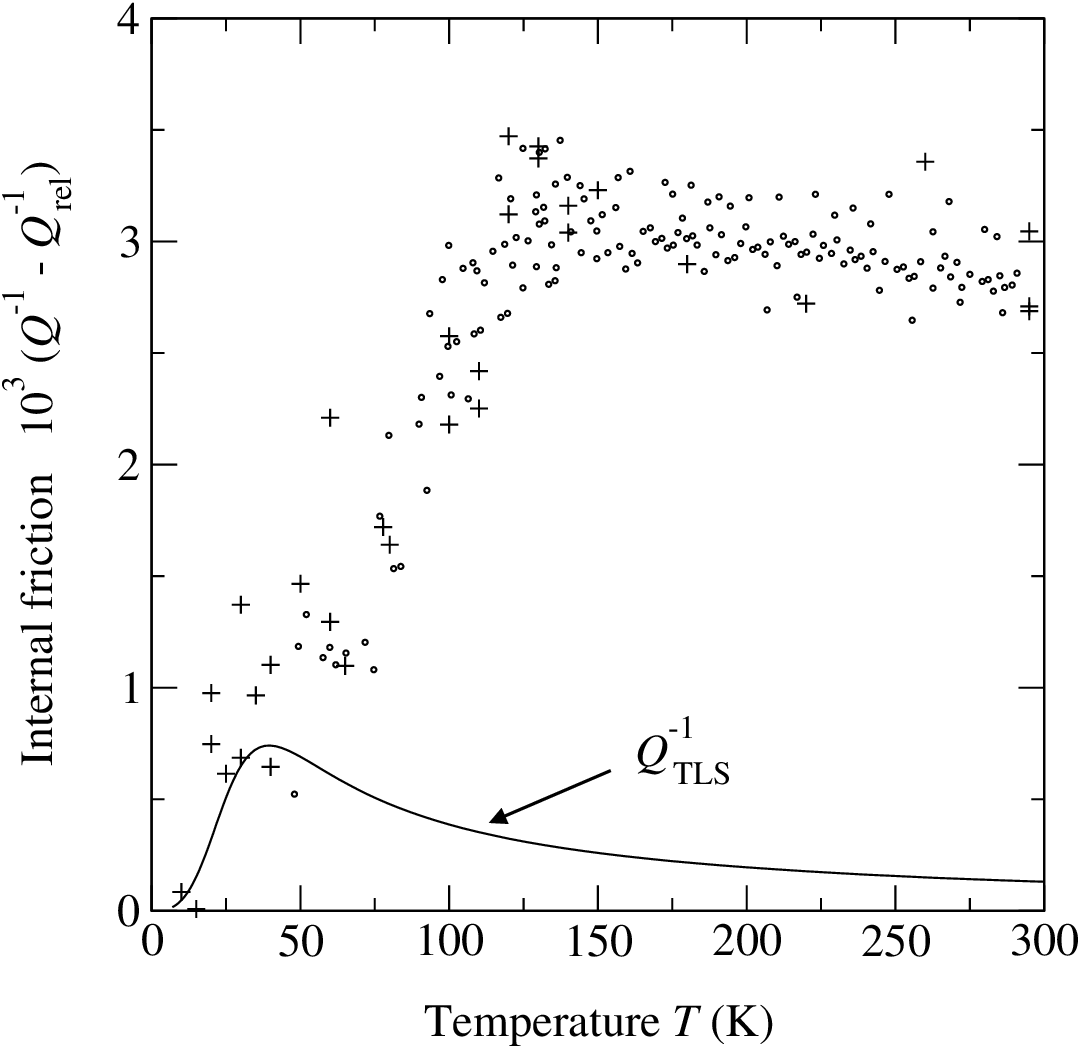}
\caption{The internal friction observed in Brillouin scattering after subtraction of the TAR contribution shown in Fig.$\; (3b)$.
The crosses are from \cite{Vac76} while the dots are from \cite{Tie92}. The solid line is a rough estimate for the small tunneling
contribution extrapolated from data presented in \cite{Top96}.}
\end{figure}
At these hypersonic frequencies and temperatures, one can anticipate a contribution $Q_{\rm TLS}^{-1}$ arising from
the relaxation of two-level systems (TLS) by incoherent tunneling.
Generally, $Q_{\rm TLS}^{-1}(T)$ exhibits a plateau separating a regime $\Omega \tau _ {\rm min} \gg 1$ at
low $T < \stackrel{\textstyle \sim}{T} $ from
$\Omega \tau _ {\rm min} \ll 1$ at higher $T$, as shown in \cite{Rau95} where $\tau _ {\rm min}$ and
$\stackrel{\textstyle \sim}{T} \propto \Omega ^{1/3}$  are defined.
The position, extent, and height of this plateau are $\Omega$-dependent, as emphasized in Fig.~6 of \cite{Top96}.
According to \cite{Rau95}, the extent of the plateau should shrink to zero at sufficiently high frequencies such that
$\stackrel{\textstyle \sim}{T} > T^*$, where $T^*$ is a frequency independent characteristic temperature marking
the end of the plateau region.
In $v$-SiO$_2$, it has been observed that $T^* \simeq 5$~K \cite{Tie92}.
The plateau becomes then a hump slightly above $\stackrel{\textstyle \sim}{T}$.
As shown in \cite{Top96}, for $\nu = 46$~MHz one should have $\stackrel{\textstyle \sim}{T} \simeq 3$~K.
This value increases with $\Omega ^{1/3}$, as confirmed from sound-damping data between 330 and 930 MHz \cite{Jon64},
showing a shoulder on the low $T$ side of the TAR peak.
Hence, at 35 GHz one expects $\stackrel{\textstyle \sim}{T} \sim 30$~K, giving a broad TLS hump centered around 40 K.
$Q_{\rm TLS}^{-1} \propto T^3$ below this hump, and $Q_{\rm TLS}^{-1} \propto T^{-1}$ above it \cite{Rau95}.
Extrapolating the observations reported in \cite{Top96} we posit
$$Q_{\rm TLS}^{-1} \; = \; \frac{1.3 \times 10^{-3}}{(30/T)^3 \, + \, (T/30)} \; \; \eqno{(12)}$$
as a rough estimate for this relatively small contribution to $Q^{-1}$.
This curve, which peaks around 40~K, is shown in Fig.~5. The difference
$$Q_{\rm anh}^{-1} \; = \; Q^{-1} \, - \, Q_{\rm rel}^{-1} \, - \, Q_{\rm TLS}^{-1} \; \; \eqno{(13)}$$
will be used below to analyse the anharmonic damping.
Since there seems to be a small tunneling contribution $Q_{\rm TLS}^{-1}$, then there ought to be by
the Kramers-Kronig relation a corresponding contribution $(\delta v)_{\rm TLS}$.
As shown in Eq.~(2.96) of \cite{Rau95}, this should produce at high-$T$ a frequency dependent term in $T \, {\rm ln} \, \Omega$.
As shown in Appendix B, there are reasons to believe that it is this term that produces the difference in slopes
indicated by short dashed lines in Fig.~4.

To conclude, we find a relatively large $Q_{\rm anh}^{-1}$ at hypersonic frequencies.
It is of the same order of magnitude as $Q_{\rm rel}^{-1}$.
The small additional term $Q_{\rm TLS}^{-1}$ that has been discussed above is a correction of minor importance compared
to $Q_{\rm anh}^{-1}$, as obvious from Fig.~5.

\section{The thermal relaxation time}

In the spirit of (I) we now analyze the anharmonic damping with the expression
$$Q_{\rm anh}^{-1} \; = \; A \; \frac{\Omega \tau_{\rm th}}{1 + \Omega ^2 \tau_{\rm th}^2} \; \; , \eqno{(14)}$$
where $\tau_{\rm th}$ is the mean lifetime of the thermal modes and
$A(T) = \gamma^2 C_{\rm v} T v / 2 \rho v_{\rm D}^3$
is the prefactor given in Eq.$\; (5b)$ of (I),
with $C_{\rm v}$ the specific heat per unit volume and $v_{\rm D}$ the Debye velocity.
We remark that all quantities entering $A(T)$ are known, except for the mean-square average Gr\"uneisen parameter $\gamma^2$.
Thus we can directly plot the quantity $\gamma^2 Q_{\rm anh}^{-1} /A$ as it is independent from $\gamma^2$.
The result obtained using the values $Q_{\rm anh}^{-1}$ taken from Fig.~5 is drawn in Fig.$\; (6a)$.
\begin{figure}
\includegraphics[width=8.5cm]{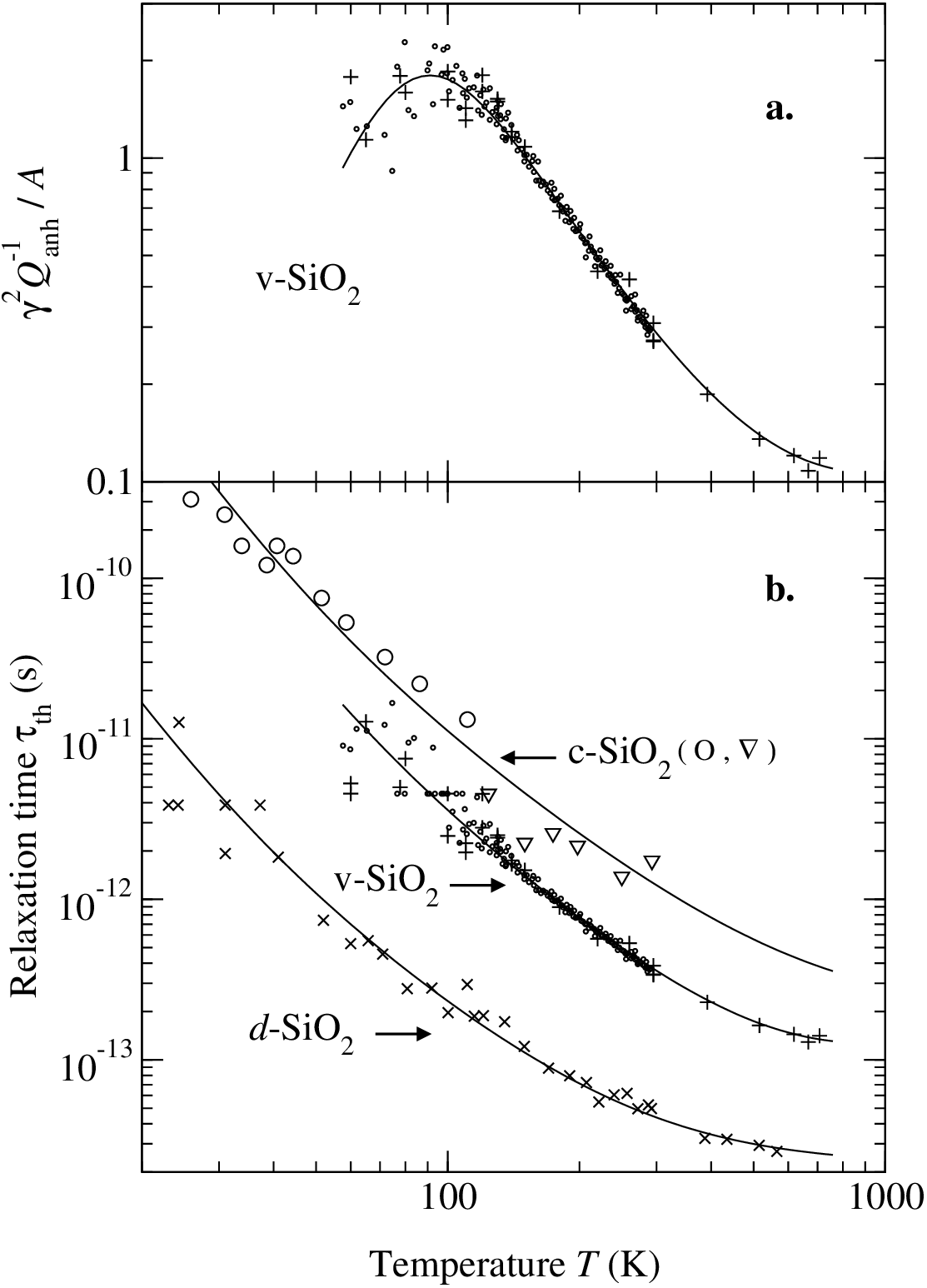}
\caption{{\bf a}: The anharmonic damping contribution scaled by $A(T)/\gamma^2$.
The solid line is a guide to the eye calculated from the corresponding line in b. The peak is located at $T_{\text Max}$ (see Appendix C) and its
height gives $\gamma^2/2$.
{\bf b}: The values of $\tau_{\rm th}$ for $v$-SiO$_2$ obtained from the data in a after solving Eq. (14). Data for x-cut crystal quartz and
$d$-SiO$_2$ from (I) are shown for comparison. The lines for  $v$-SiO$_2$ and  $d$-SiO$_2$  are guides to the eye which are sums of
negative powers of $T$, while the line for $c$-SiO$_2$ is calculated from Eq. (15).}
\end{figure}
Points below 60~K are not shown as the data is too noisy at low $T$ to be significant after division 
by $A \propto C_{\rm v} T$.
This plot exhibits a peak for $\Omega \tau_{\rm th} \, = \, 1$.
From the height of this peak one reads $\gamma^2 \simeq 3.6$.
With this determination, Eq. (14) can then be solved for $\Omega \tau_{\rm th}$, and thus for $\tau_{\rm th}$.
The procedure is explained in Appendix C.
The result is shown in Fig.$\; (6b)$.
One observes that $\tau_{\rm th}(T)$ is proportional to $T^{-2}$ over the main part of the data, from
$\sim 100$~K to  $\sim 300$~K.
At higher $T$, $\tau_{\rm th}$ tapers off, while at lower $T$ there is a hint that $\tau_{\rm th}(T)$ might diverge
faster than in $T^{-2}$.
The solid line is a guide to the eye which is a sum of powers of $T$.
The same law is then used to trace the guide to the eye in Fig.$\; (6a)$.

In Fig.$\; (6b)$ we have traced for comparison the values of the relaxation time of the thermal modes for two other cases.
Firstly, we observe that $\tau_{\rm th}(T)$ obtained in (I) for densified silica glass, $d$-SiO$_2$, 
is substantially shorter than in $v$-SiO$_2$. However, it shows a very similar functional dependence on $T$. 
Secondly, in order to deepen the understanding, we applied the same method of analysis to data on the LA-mode of X-cut crystal quartz, 
$c$-SiO$_2$.
We used to this effect the attenuation coefficient measured by B\"ommel and Dransfeld at 1~GHz \cite{Boe60}.
At temperatures above 150~K, we completed these data using the width of the Brillouin line reported in \cite{Bon94}.
These data were analyzed similarly to Fig.$\; (6a)$ leading to $\gamma^2 \simeq 0.95$.
The corresponding  $\tau_{\rm th}(T)$ is shown in Fig.$\; (6b)$.
In this case, the line through these points is {\em not} a guide to the eye as for the two glasses, but it is an independent
determination of $\tau_{\rm th}(T)$ using the well-known kinetic expression for the thermal conductivity,
$$\kappa \; = \; {\scriptstyle \frac{1}{3}} \; C_{\rm v} v_{\rm D}^2 \tau_{\rm th} \; \; . \eqno{(15)}$$
We observe that in this case the value of $\tau_{\rm th}$ derived from $\kappa$ (line) is in remarkable agreement with that obtained from the
measurement of anharmonic damping of the acoustic modes.
This emphasizes the physical significance of $\tau_{\rm th}$: it really is the mean lifetime of the excitations in the
thermal bath which at sufficiently high $T$ is entirelly controlled by Umklapp (U) processes.
As pointed out in \cite{Zim60}, this produces a relaxation proportional to the phonon population, and thus at high $T$
one has $\tau_{\rm th} \propto T^{-1}$.
At intermediate $T$, where $C_{\rm v} \; \stackrel{\textstyle \sim}{\propto} \; T$ and
$\kappa \; \stackrel{\textstyle \sim}{\propto} \; T^{-1}$, Eq. (15) indicates that
$\tau_{\rm th} \propto T^{-2}$.
This is indeed observed in $c$-SiO$_2$ over a large range of $T$, as seen in Fig.$\; (6b)$.

Equation~(15) assumes propagating thermal phonons.
For this reason, it becomes invalid in glasses at soon as $T$
increases beyond the thermal conductivity plateau located around 10~K in $v$-SiO$_2$.
However, we remark that the functional dependence $\tau_{\rm th} (T)$ observed in the two glasses is very similar in shape
to that in $c$-SiO$_2$.
We have no simple explanation for this.
In glasses, there are two competing effects that modify the picture presented above.
On the one hand, the strict quasi-momentum conservation which is invoqued in U-processes is strongly relaxed, and this
must greatly enhance the interactions in the thermal bath, decreasing $\tau_{\rm th}$.
This could account for the observed difference between crystal and glasses.
On the other hand, the thermal modes are not expected to be propagating plane-waves but they are at best diffusive.
This restricts the spatial extent of the modes, greatly decreasing their overlap and thereby their interactions,
which increases $\tau_{\rm th}$.
We believe it is the latter effect which produces the much longer $\tau_{\rm th}$ in $v$-SiO$_2$ compared to $d$-SiO$_2$.
Indeed, in $d$-SiO$_2$ the boson peak is strongly reduced \cite{Ina99}.
This increases the crossover frequency $\omega _{\rm co}$
beyond which the acoustic excitations become diffusive, as recently confirmed by inelastic x-ray scattering \cite{Ruf03}.
This in turn increases the mean spatial extent of the modes and thereby their interactions leading to a faster thermalization.

Finally, we remark that the above analysis gives values for the mean Gr\"uneisen parameter $\gamma^2$.
We find 0.95 for the LA-mode of x-cut quartz, 3.6 for $v$-SiO$_2$, and 8 for $d$-SiO$_2$.
In the case of crystal quartz, the agreement between the line calculated from the thermal conductivity $\kappa$
and that obtained from $Q^{-1}$ gives a solid support for the value of $\gamma^2$.
Unfortunately, one cannot make a similar comparison for the glasses.

\section{Summing-up}

\subsection{The internal friction}
Figure~7 summarizes the present analysis of the internal friction observed in $v$-SiO$_2$ with Brillouin scattering around 35 GHz.
One identifies two main contributions: $Q_{\rm rel}^{-1}$ arising from thermally activated relaxation and
$Q_{\rm anh}^{-1} = Q^{-1} - (Q_{\rm rel}^{-1} + Q_{\rm TLS}^{-1})$ which results from network viscosity.
Around room $T$, $Q_{\rm anh}^{-1}$ is nearly twice as large as $Q_{\rm rel}^{-1}$ .
The incoherent tunneling contribution $Q_{\rm TLS}^{-1}$, if it really exists at these high frequency and temperatures,
mainly produces the small hump around 40 K.
Although there is a hint for such a feature in the data shown in Fig.~7, it would need to be confirmed by
more precise measurements.
Also, we neglected in this analysis a possible contribution arising from the quasi-harmonic oscillators of the
soft-potential model \cite{Par94}.
This is explained in Appendix D.
\begin{figure}
\includegraphics[width=8.5cm]{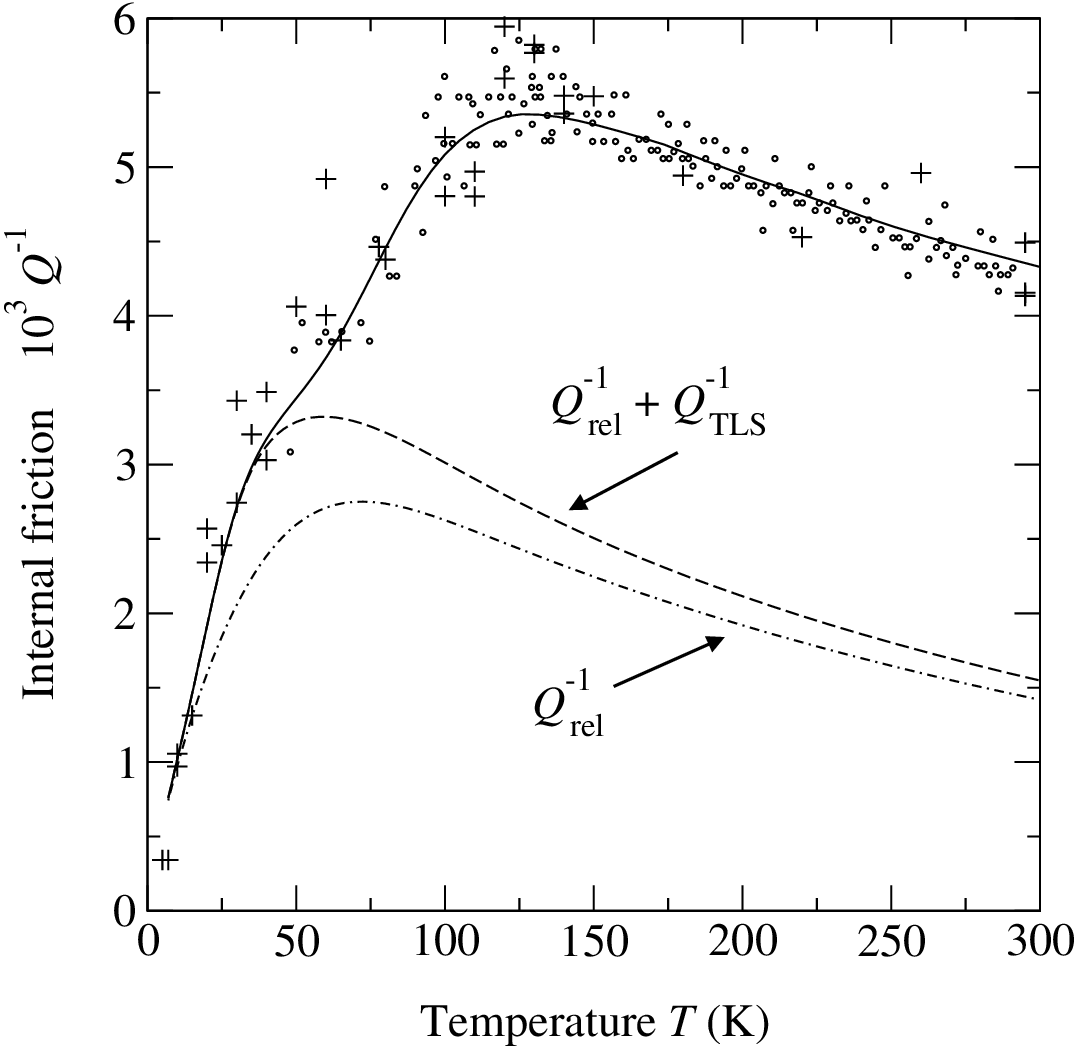}
\caption{The internal friction at 35 GHz from the Brillouin scattering measurements of \cite{Tie92} ($\circ$) and \cite{Vac76} (+).
The solid curve is the entire adjustment made here which is the sum of three contributions $Q^{-1}_{\rm rel}$, $Q^{-1}_{\rm TLS}$,
and $Q^{-1}_{\rm anh}$.}
\end{figure}

As $\nu$ is increased beyond 35 GHz, we find that the value $Q_{\rm rel}^{-1}$ calculated with Eq.~(9a) saturates.
This saturation essentially results from the cut-off in $f(\Delta)$.
For example at room $T$, $Q_{\rm rel}^{-1}$ does not increase beyond
$\sim$1.5$\times$10$^{-3}$.
On the other hand, at sufficiently high $T$, $Q_{\rm anh}^{-1}$ continues to grow with $\Omega$ according to
$Q_{\rm anh}^{-1} = A \Omega \tau_{\rm th}$ which applies as long as $\Omega \tau_{\rm th} \ll 1$.
We also find that $A \tau_{\rm th}$ is practically constant from 100 to 300~K.
For these reasons, the total $Q^{-1}$ becomes dominated by network viscosity, it continues to increase proportionally to
$\Omega$ up to several hundred GHz, and it is nearly $T$-independent from 100~K to 300~K.
This is in excellent agreement with the results on acoustic propagation reported for $v$-SiO$_2$ thin films by
Zhu, Maris, and Tauc \cite{Zhu91}.
These authors find that the mean free path, $\ell ^{-1}$, grows in $\Omega ^2$ from $\sim$30 GHz to $\sim$300 GHz,
and that it is nearly independent of $T$ from 80 to 300~K.
Since $\ell ^{-1} \equiv Q^{-1} \Omega / v$, the dominance of $Q_{\rm anh}^{-1}$, its growth $\propto \Omega$, and its near
constancy in $T$, fully account for the data reported in \cite{Zhu91}.
Our predictions are also fully consistent with three data points measured with UV Brillouin scattering from $\sim$50 to
$\sim$100 GHz \cite{Mas04,Ben05}.

On the other hand, at ultrasonic frequencies, {\em i.e.} much below 35 GHz, the contribution of $Q_{\rm anh}^{-1}$ 
which decreases proportionally to $\Omega$ becomes completely negligible compared to $Q_{\rm rel}^{-1}$.
This justifies the analysis based only on TAR that was performed in relation with Fig.~1.

\subsection{The velocity changes}
\begin{figure}
\includegraphics[width=8.5cm]{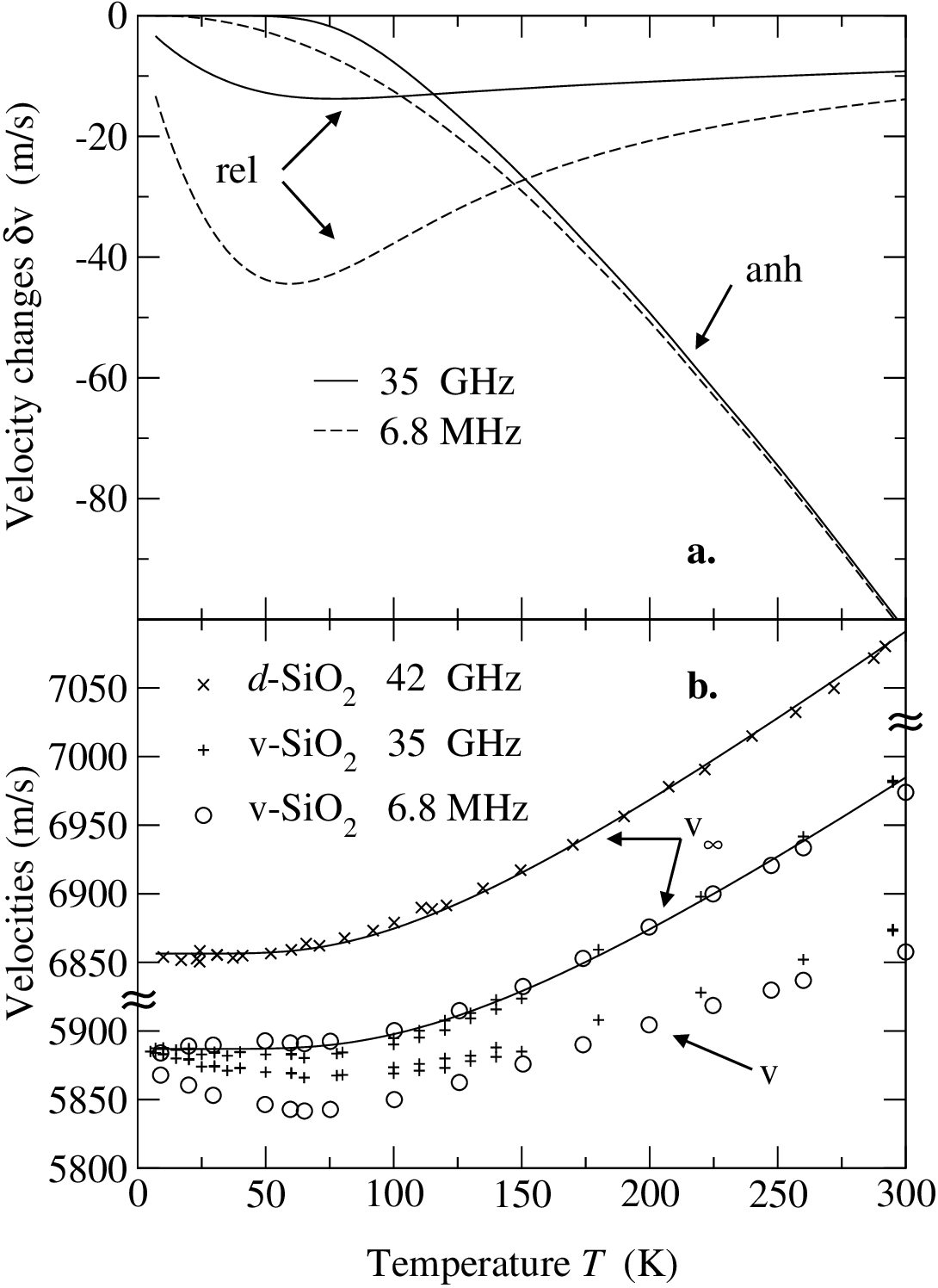}
\caption{{\bf a}: Calculated velocity changes produced by TAR (rel) and by network viscosity (anh) at two typical frequencies.
{\bf b}: The measured velocities at the same frequencies (lower data points) and the unrelaxed
values $v_\infty$ after correction for TAR and network viscosity (upper data points).
The line is a fit of $v_\infty$ to Wachtman's equation as a guide to the eye.
The $v_\infty$ data for $d$-SiO$_2$ from (I) are also shown for comparison.}
\end{figure}
Figure$\; (8a)$ shows the velocity changes calculated from Eq.$\; (9b)$ for TAR, and using
$$\left( \frac {\delta v}{v} \right)_{\rm anh} \; = \; -\frac {A}{2} \; 
\frac {1}{1 + \Omega ^2 \tau_{\rm th} ^2} \; \;  \eqno{(16)}$$
for the network viscosity, this at two typical frequencies.
One notices that the increase in $\nu$ from 6.8~MHz to 35~GHz leads to a large reduction of $| \delta v_{\rm rel}|$.
On the other hand, $| \delta v_{\rm anh}|$, although quite large above 100~K, changes relatively little with $\nu$.
In particular, the dashed curve is entirely in the regime $\Omega \tau_{\rm th} \ll 1$, so that it is practically independent of
$\Omega$ at lower $\Omega$-values.
This justifies the approach used to determine ${\cal C}$ in Fig.~2 where the anharmonicity contribution was simply neglected.
At sufficiently high-$T$, anharmonicity always leads to a quasi-linear decrease of $v$ with $T$, very similar
to what is observed in crystals \cite{Lan66,Gar75}.
This has been seen in ultrasonic measurements in a large number of glasses, {\em e.g.} in \cite{Cla78}
where this effect is interpreted as such.
At these frequencies, one is in a regime where the velocity changes are dominated by network viscosity while the damping mainly
originates from TAR.
It is generally necessary to go up to Brillouin frequencies to observe in $Q^{-1}$ both contributions simultaneously.

The velocity changes in tetrahedrally coordinated glasses, in particular in $v$-SiO$_2$ and in $d$-SiO$_2$,
are more complex than described above.
Fig.$\; (8b)$ shows the observed velocities, $v$, and the unrelaxed value that is corrected for both TAR and
network-viscosity dispersions, $v_{\infty} = v - [(\delta v)_{\rm rel} + (\delta v)_{\rm anh}]$. 
As observed, the experimental points collapse to a $v_{\infty}$ which on this scale is essentially $\Omega$-independent.
However, rather than being constant, $v_{\infty}$ increases considerably with $T$.
In the same figure, the results obtained in (I) for $d$-SiO$_2$ are shown for comparison.
Both solid lines are guides to the eye obtained by an {\em ad hoc} adjustment to Wachtman's equation \cite{And66}.
The behaviour of $v_{\infty}$ is remarkably similar in both glasses,
pointing to an anomalous hardening of silica with increasing $T$.
As discussed in (I), the earlier proposal that this is a manifestation of large structural inhomogeneities \cite{Kul74}
is not supported by observations.
The anomalous hardening has recently been attributed to a progressive local polyamorphic transition
associated with abrupt rotations of randomly distributed Si-O-Si bonds, found in simulations \cite{Hua04a}.
Thermal agitation would redistribute the Si-O-Si bond angles in configurations where they are more resilient.
However, contrary to \cite{Hua04b}, our results indicate that a very similar mechanism would then be active in
permanently densified silica.
Repeating the warning made in (I), although $v_{\infty}$ has a clear physical significance,
it would be nearly impossible to perform at the moment an experiment to directly measure it. 
This is even more so in $v$-SiO$_2$ than in $d$-SiO$_2$, owing to the lower $\omega_{\rm co}$ of the former glass.

\subsection{The tunneling strength}
It remains to compare the value ${\cal C} = 1.4 \times 10^{-3}$ found in Section III to the tunneling-strength parameter
$C \simeq 3.0 \times 10^{-4}$ from the literature.
To do this, both values are inserted in Eq.~(11) to calculate the cut-off energy $W$ that applies to the present case.
We find $W \simeq 7$~K, as opposed to the TLS value $W \simeq 4$~K  derived from the
low-$T$ specific heat data \cite{Buc92}.
Such a larger value of $W$ was already reported, resulting from a similar analysis in \cite{Kei93}.
Its origin was convincingly explained there as arising from a different weighing of the distribution of $W$-values,
in $W^{-4}$ for the specific heat, and in $W^{-1}$ for the ultrasonic absorption.
The value of $\cal C$ is thus fully in line with expectations.

\section{Discussion}

We have shown that ultrasonic and hypersonic damping and dispersion in vitreous silica at temperatures above the
quantum regime originate in at least two main processes: thermally activated relaxation and network viscosity.
These have different $\Omega$ and $T$ dependences which leads to a non-trivial variation of the scattering
linewidth $\Gamma$ with $\Omega$.
At a given $T$, TAR dominates the damping at low $\Omega$, while anharmonicity dominates at sufficiently high $\Omega$.
Between these two regimes there occurs a ``dynamical crossover'' which for $v$-SiO$_2$ at room $T$ falls near our Brillouin
frequency, as shown in Fig.~10 of \cite{Zhu91} and confirmed in Fig.~7 above.
Above this crossover, the damping is dominated by $Q_{\rm anh}^{-1} = A \Omega \tau_{\rm th}$,
which implies that $\Gamma \propto \Omega^2$.
As $\Omega$ is increased far beyond optical Brillouin frequencies, another broadening 
mechanism eventually comes into play.
Around 300~GHz, one expects the onset of a new contribution to $Q^{-1}$, growing with a higher power of $\Omega$,
possibly in $\Omega^3$, as suggested in \cite{Vac97}.
This additional contribution is currently thought to result from the hybridization of the acoustic modes with
boson-peak excitations \cite{Gur03,Ruf05}, as inelastic x-ray scattering observations on $d$-SiO$_2$
strongly suggest \cite{Ruf03,Cou03}.
Unfortunately, inelastic x-ray scattering at sufficiently small scattering vectors and with sufficiently high
energy resolution could not yet be performed on $v$-SiO$_2$ to settle the issue of this onset.
This hybridization eventually leads to the Ioffe-Regel crossover at $\omega _{\rm co}$,
above which the excitations have lost their plane-wave character.
Then, the wave-vector $q$ does not remain a good label for the modes which are at best diffusive \cite{Fab96}.
The excitations of the thermal bath are expected to be of that nature in glasses, as discussed in Section V.

Power laws $\Gamma \propto \Omega^n$, or $\Gamma \propto q^n$ which is equivalent as long as the frequency is well below
the Ioffe-Regel limit so that $q \propto \Omega$, are often employed to represent the damping data $\Gamma(\Omega)$.
As just explained, $\Gamma \propto \Omega^2$ is a reasonable approximation in the region between the dynamical
crossover and the Ioffe-Regel one.
On the opposite, such laws should be viewed as quite rough approximations in the TAR-dominated regime.
The reason is that the dependence of $\Gamma$ on $\Omega$ predicted by Eq. 9a is more complicated.
It cannot be written simply in terms of a $T$-dependent exponent $n(T)$.
For example, in $v$-SiO$_2$ at room $T$, a value $n=1.3$ is given in \cite{Sco03}, while $n=1.8$ is
derived from \cite{Wie00}.
We conclude that power laws in the TAR-dominated regime are {\em ad hoc} devices which only represent approximately
the data, especially if the latter are sufficiently noisy to mask the disagreements.

At this point, it is legitimate to ask to what extent anharmonicity might be important to sound damping in 
other glasses.
The answer to this question depends on two factors:
i) the strength, density, and distribution of the relaxing defects;
ii) the strength of the anharmonicity.
It is conceivable that in glasses that contain a large quantity of defects up to high values of $V_0$ and 
$\Delta_{\rm C}$, TAR would mask the anharmonicity.
In that case the dynamical crossover might move up to nearly $\omega_{\rm co}$, so that a regime
$\Gamma \propto \Omega^2$ might not be observable at all.
We have found such a case in lithium diborate, $v$-Li$_2$B$_4$O$_7$ \cite{Ruf05}.
This type of situation is expected for most polymers in which the tunneling strength $C$ is 
quite large \cite{Poh02}.
On the other hand, there exist many glasses or amorphous materials for which $C$ is quite small,
well below $10^{-4}$ \cite{Poh02}.
Such are the amorphous group IV semiconductors, C, Si, and Ge.
In that case, the anharmonicity is expected to dominate over a considerable range of $\Omega$
and a calculation to that effect has already been performed \cite{Fab99}.
Of the tetrahedrally coordinated glasses, it will be particularly interesting to reinvestigate
$v$-GeO$_2$ and $v$-BeF$_2$.
Difficulties met in obtaining a quantitative description of damping and dispersion in 
germania on the basis of TAR alone, described in \cite{Her98}, might find their resolution by including in the analysis
the cut-off $\Delta_{\rm C}$ and the network-viscosity contribution.

To summarize, TAR essentially explains acoustic damping results in silica glass up to a few GHz.
At higher frequencies, the anharmonic coupling of hypersound to the modes that form the entire thermal bath becomes
progressively dominant.
This conclusion, in line with previous observations \cite{Vac81} and simulations \cite{Fab99}, corrects
statements to the contrary often found in the literature, {\em e.g.} in \cite{Wie00,Wie01,Sur04}.
The anharmonicity which in our view produces $Q^{-1}_{\rm anh}$ is a coupling of sound to the whole bath, in addition to
the relatively small fraction of modes forming the boson peak.
This process, which we call network viscosity, is in a way quite similar to the usual lattice viscosity of crystals.

\appendix

\section{Estimating Eqs. (9)}

For the Marquardt-Levenberg adjustment of Eqs. (9), it is convenient to use the parameters
$\cal C$, $V_0$, log$_{10} \, \tau _0$, $\zeta$, and $\delta$. Taking as integration variable $x \equiv V/V_0$,
and defining the auxiliary variables $\alpha \equiv V_0/T$ and $y \equiv \Omega \tau_0$, Eq.$\; (9a)$ becomes
\begin{widetext}
$$Q_{\rm rel}^{-1} \; = \; {\cal C} \: \Phi ( \sqrt{2} \, \delta / \alpha ) \: \alpha \: \int_{0}^{\infty}
x^{- \zeta} {\rm exp} ( - {\scriptstyle \frac{1}{2}} \, x^2 )
\; \frac { y \: {\rm exp}(\alpha x)}{1 + y^2 \: {\rm exp} (2\alpha x)} \: dx \; \;  . \eqno{({\rm A}1)}$$
\end{widetext}
Good results are obtained with a simple trapezoidal integration, taking a geometric progression for the vector of $x$-values.
We used $x$ starting at $\simeq 10^{-4}$ and ending at $\simeq 10$, in steps of 1\%.
Eq.$\; (9b)$ is handled similarly.

\section{Velocity changes due to incoherent tunneling}

One observes on the upper curves of Fig.~4 that there is a small difference in the high temperature slopes for the
two different frequencies.
The origin of this effect might be in the velocity dispersion associated, by the Kramers-Kronig transform, with the
relatively small damping $Q^{-1}_{\rm TLS}$  produced by incoherent tunneling.
At high $T$, the damping is given by the third equation (2.94) of \cite{Rau95},
$$Q^{-1}_{\rm TLS} \; = \; \frac{\pi \, C}{4 \, r_{\rm min}} \; \frac{\hbar \Omega}{k_{\rm B} T} \; \; . \eqno{({\rm B}1)}$$
The corresponding $\delta v _{\rm TLS}$ is derived from the third equation (2.96) of \cite{Rau95},
which gives for the $\Omega$-dependent part
$$\delta v_{\rm TLS} \; \simeq \; 0.45 \, \frac{C}{T^*} \; ({\rm ln}\, \Omega) \, T \; \; . \eqno{({\rm B}2)}$$
The temperature $T^*$ marks the end of the low-$T$ plateau region in $Q^{-1}$ and it is given by Eqs. (2.78) and (2.79) of \cite{Rau95}.
Owing to the logarithmic dependence in (B2), it is the difference of slopes at two frequencies which is significant,
\begin{widetext}
$$\frac{\Delta (\delta v_{\rm TLS})}{T} \, \equiv \,  \frac {(\delta v_1)_{\rm TLS} - (\delta v_2)_{\rm TLS}}{T}  \simeq 
\, 0.45 \, \frac{C}{T^*} \, {\rm ln}\,(\Omega_1 / \Omega_2) \; . \eqno{({\rm B}3)}$$
\end{widetext}
With the values in Fig.~4, and calculating $T^*$ following \cite{Rau95}, one finds
$C \approx 7 \times 10^{-5}$. As expected, this is smaller than the tunneling strength
found from specific heat measurements at low temperatures \cite{Poh02}, the reduction being of the order
of $\Delta _{\rm C}/T$.
Hence, it is not unreasonable to assign the difference in slopes observed at high-$T$ in Fig. 4 to the Kramers-Kronig
transform of $Q^{-1}_{\rm TLS}$.\\

\section{Solving Eq. (14) for the relaxation time}

It is in principle trivial to solve the quadratic equation $y=x/(1+x^2)$ for $ x \equiv \Omega \tau_{\rm th}$ in terms
of the measured $y \equiv Q^{-1}/A$.
However, imaginary roots do occur in the region around $x=1$ owing to fluctuations
in the data leading to some points for which $y>1/2$.
These are the points whose ordinate lies above the maximum of the solid curve drawn
in Fig.$\; (6a)$.
For these data points, we simply replaced $x$ by 1, as the nearest real solution.
This produces a line of data points with $\tau_{\rm th} = 1/\Omega$ in the presentation of Fig.$\; (6b)$.
For the data points with $y<1/2$, one must select between the upper and the lower root of the quadratic equation.
This is set by the location of the maximum in the solid curve, $T_{\rm Max}$.
For $T < T_{\rm Max}$, the upper root applies since $\Omega \tau_{\rm th} > 1$, while
the lower root applies at higher temperatures.\\

\section{Another possible source of damping}

We have not considered the damping
arising from the weak anharmonicity of the nearly harmonic oscillators (HO) of the soft
potential model in the domain $\eta > 0$ \cite{Par94}.
This contribution should be most active at ``intermediate'' frequencies.
In $v$-SiO$_2$, its strength has actually been predicted to be largest near our Brillouin frequency \cite{Par94}.
It must be noted that the entire $(\Omega,T)$-domain covered by the propagating pulse
measurements of \cite{Zhu91} falls within this  ``intermediate'' frequency range
where the additional damping obeys $Q^{-1}_{\rm HO} \propto C T / \sqrt{\Omega}$ \cite{Par94}.
If the strength of $Q^{-1}_{\rm HO}$ would be sufficient, one should definitely notice its effect on the
$(\Omega,T)$-dependence of the mean free
path reported in \cite{Zhu91} from 80 to 300~K and from 30 to 300~GHz.
On the contrary, the mean free path is found practically $T$-independent and it increases $\propto \Omega^2$,
in agreement with the network-viscosity contribution.
It seems thus justified to neglect $Q^{-1}_{\rm HO}$ in our analysis of the Brillouin results up to 300~K.
The difficulty in estimating the size of $Q^{-1}_{\rm HO}$ apparently lies in finding
the proper value for the tunneling-strength parameter $C$ that applies to it.
The value $C=$ 3.0$\times$10$^{-4}$, appropriate for TLS at very low temperatures \cite{Poh02}, is certainly too large here.
Indeed, using this value in Eq. (3.18) of \cite{Par94}, we calculate $Q^{-1}_{\rm HO} \simeq$ 5$\times$10$^{-5} \times T$(K),
which at 300~K is larger than the entire $Q^{-1}$ observed.
>From an analysis of heat-release measurements there are grounds to adopt here a value of $C$ which is
at least five times smaller \cite{Par93}.
Even so, the resulting $Q^{-1}_{\rm HO}$ is still too large to agree with the results of \cite{Zhu91}.
Our Brillouin measurements up to 700~K shown in Fig.$\; (3b)$ suggest that there could be a small contribution
$\propto T$ that starts being felt above $\sim 400$~K.
This contribution would then be of the order of $10^{-6} \times T$(K), which implies a very small effective $C$
for this particular damping mechanism.\\


\begin{references}

\bibitem{And55}O.L. Anderson and H.E. B\"ommel, J. Am. Ceram. Soc.
{\bf 38}, 125 (1955).

\bibitem{And72}P.W. Anderson, B.I. Halperin, and C.M. Varma, Phil. Mag.
{\bf 25}, 1 (1972).

\bibitem{Phi72}W.A. Phillips, J. Low Temp. Phys. {\bf 7}, 351 (1972).

\bibitem{Jae72}J. J\"{a}ckle, Z. Physik {\bf 257}, 212 (1972).

\bibitem{Jae76}J. J\"ackle, L. Pich\'e, W. Arnold, and S. Hunklinger,
J. Non-Cryst. Solids {\bf 20}, 365 (1976).

\bibitem{Gil81}K.S. Gilroy and W.A. Phillips, Phil. Mag. {\bf 43}, 735 (1981).

\bibitem{Wie00}J. Wiedersich, S.V. Adichtchev, and E. R\"ossler,
Phys. Rev. Lett. {\bf 84}, 2718 (2000).

\bibitem{Wie01}J. Wiedersich, N.V. Surotsev, V.N. Novikov, E. R\"ossler, and A.P. Sokolov,
Phys. Rev. B {\bf 64}, 064207 (2001).

\bibitem{Sur04}N.V. Surotsev, V.K. Malinovsky, Yu.N. Pal'yanov, A.A Kalinin, and A.P. Shebanin,
J. Phys.: Condens. Matter {\bf 16}, 3035 (2004).

\bibitem{Her98}J. Hertling, S. Bae{\ss}ler, S. Rau, G. Kasper, and S. Hunklinger,
J. Non-Cryst. Solids {\bf 226}, 129 (1998).

\bibitem{Phi87}W.A. Phillips, Rep. Prog. Phys. {\bf 50}, 1657 (1987).

\bibitem{Tie92}D. Tielb\"urger, R. Merz, R. Ehrenfels, and S. Hunklinger,
Phys. Rev. B {\bf 45}, 2750 (1992).

\bibitem{Poh81}R.O. Pohl, in {\em Amorphous Solids: Low-Temperature Properties}, W.A. Phillips Ed. (Springer, Berlin, 1981)
pp. 27-52.

\bibitem{Kar83}V.G. Karpov, M.I. Klinger, and F.N. Ignat'ev, Sov. Phys. JETP {\bf 57}, 439 (1983).

\bibitem{Ili87}M.A. Il'in, V.G. Karpov, and D.A. Parshin, Sov. Phys. JETP {\bf 65}, 165 (1987).

\bibitem{Buc92}U. Buchenau, Yu.M. Galperin, V.L. Gurevich, D.A. Parshin, M.A. Ramos, and H.R. Schober,
Phys. Rev. B {\bf 46}, 2798 (1992).

\bibitem{Par94}D.A. Parshin, Phys. Solid State {\bf 36}, 991 (1994).

\bibitem{Kei93}R. Keil, G. Kasper, and S. Hunklinger,
J. Non-Cryst. Solids {\bf 164-166}, 1183 (1993).

\bibitem{Abr70}M. Abramowitz and I.A. Stegun, {\em Handbook of Mathematical Functions}
(Dover, New York, 7th Printing 1970), Ch. 19, particularly p. 687.

\bibitem{Gil93}L. Gil, M.A. Ramos, A. Bringer, and U. Buchenau,
Phys. Rev. Lett. {\bf 70}, 182 (1993).

\bibitem{Bon91}J.P. Bonnet,
J. Non-Cryst. Solids {\bf 127}, 227 (1991).

\bibitem{Poh02}R.O. Pohl, Xiao Liu, and EJ. Thompson, Rev. Mod. Phys.
{\bf 74}, 991 (2002).

\bibitem{Vac81}R. Vacher, J. Pelous, F. Plicque, and A. Zarembowitch, 
J. Non-Cryst. Solids {\bf 45}, 397 (1981).

\bibitem{Deb32}P. Debye and F.W. Sears, Proc. Natl. Acad. Sci. USA, {\bf 18}, 409 (1932).

\bibitem{Str64}R.E. Strakna and H.T. Savage, J. Appl. Phys. {\bf 35}, 1445 (1964).

\bibitem{Gil80}K.S. Gilroy and W.A. Phillips, J. Non-Cryst. Solids {\bf 35\&36}, 1135 (1980).

\bibitem{Hun76}S. Hunklinger and W. Arnold, in {\em Physical Acoustics},
Vol. XII, edited by W.P. Mason and R.N. Thurston (Academic Press, New York, 1976), pp. 155-215.

\bibitem{Rau95}S. Rau, C. Enss, S. Hunklinger, P. Neu, and A. W\"urger,
Phys. Rev. B {\bf 52}, 7179 (1995).

\bibitem{Vac76}R. Vacher and J. Pelous, Phys. Rev. B {\bf 14}, 823 (1976).

\bibitem{Pel76}J. Pelous and R. Vacher, Solid State Commun. {\bf 18}, 657 (1976).

\bibitem{Wax71}R.M. Waxler and G.W. Cleek, J. Res. Nat. Bur. Stand. (U.S.) {\bf75A}, 279 (1971).

\bibitem{Top96}K.A. Topp and D.G. Cahill, Z. Phys. B {\bf 101}, 235 (1996).

\bibitem{Jon64}C.K. Jones, P.G. Klemens, and J.A. Rayne, Phys. Lett. {\bf 8}, 31 (1964)

\bibitem{Boe60}H.E. B\"ommel and K. Dransfeld, Phys. Rev. {\bf 117},
1245 (1960).

\bibitem{Bon94}J.P. Bonnet, M. Boissier, and A. Ait Gherbi, 
J. Non-Cryst. Solids {\bf 167}, 199 (1994).

\bibitem{Zim60}J.M. Ziman Electrons and Phonons Oxford Clarendon Press 1960

\bibitem{Ina99}Y. Inamura, M. Arai, O. Yamamuro, A. Inaba, N. Kitamura, T.
Otomo, T. Matsuo, S.M. Bennington, and A.C. Hannon, Physica B 
{\bf 263 \& 264}, 299 (1999).

\bibitem{Ruf03}B. Ruffl\'e, M. Foret, E. Courtens, R. Vacher, and G. Monaco,
Phys. Rev. Lett. {\bf 90}, 095502 (2003).

\bibitem{Zhu91}T.C. Zhu, H.J. Maris, and J. Tauc, Phys. Rev. B {\bf 44}, 4281 (1991).

\bibitem{Mas04}C. Masciovecchio, A. Gessini, S. Di Fonzo, L. Comez, S.C. Santucci, and D. Fioretto,
Phys. Rev. Lett. {\bf 92}, 247401 (2004).

\bibitem{Ben05}P. Benassi, S. Caponi, R. Eramo, A. Fontana, A. Giugni, M. Nardone, M. Sampoli, and G. Viliani,
Phys. Rev. Lett. {\bf 71}, 172201 (2005).
 
\bibitem{Lan66}Landolt-B\"ornstein, New Series, Group III, Vol. 1 (Springer-Verlag, Berlin, 1966).

\bibitem{Gar75}J.A. Garber and A.V. Granato, Phys. Rev. B {\bf 11}, 3990 (1975).

\bibitem{Cla78}T.N. Claytor and R.J. Sladek, Phys. Rev. B {\bf 18}, 5842 (1978).

\bibitem{And66}O.L. Anderson, Phys. Rev. {\bf 144}, 553 (1966).

\bibitem{Kul74}M.N. Kul'bitskaya, S.V. Nemilov, and V.A. Shutilov,
Sov. Phys. Solid State, {\bf 16}, 2319 (1974).

\bibitem{Hua04a}Liping Huang and J. Kieffer, Phys. Rev. B {\bf 69}, 224203 (2004).

\bibitem{Hua04b}Liping Huang and J. Kieffer, Phys. Rev. B {\bf 69}, 224204 (2004).

\bibitem{Vac97}R. Vacher, J. Pelous, and E. Courtens, Phys. Rev. B {\bf 56}, R481 (1997).

\bibitem{Gur03}V.L. Gurevich, D.A. Parshin, and H.R. Schober, Phys. Rev. B {\bf 67}, 094203 (2003).

\bibitem{Ruf05}B. Ruffl\'e, G. Guimbreti\`ere, E. Courtens, R. Vacher, and G. Monaco, 
submitted (arXiv:cond-mat/0506287, 01 Sep 2005).

\bibitem{Cou03}E. Courtens, M. Foret, B. Hehlen, B. Ruffl\'e, and R. Vacher,
J. Phys.: Condens. Matter {\bf 15}, S1279 (2003).

\bibitem{Fab96}J. Fabian and P.B. Allen, Phys. Rev. Lett. {\bf 77}, 3839 (1996).

\bibitem{Sco03}T. Scopigno, S.N. Yannopoulos, D.Th. Kastrissios, G. Monaco, E. Pontecorvo, G. Ruocco, and F. Sette,
J. Chem. Phys. {\bf 118}, 311 (2003).

\bibitem{Fab99}J. Fabian and P.B. Allen, Phys. Rev. Lett. {\bf 79}, 1885 (1999).

\bibitem{Par93}D.A. Parshin and S. Sahling, Phys. Rev. B {\bf 47}, 5677 (1993).

\end{references}
\end{document}